# Phase separation and ferroelectric ordering in charge frustrated $LuFe_2O_{4-\delta}$

H.X. Yang, H.F. Tian, Y. Zhang, Y.B. Qin, L.J. Zeng, C. Ma,

H.L. Shi, J.B. Lu, and J.Q. Li*

*Beijing National Laboratory for Condensed Matter Physics, Institute of Physics, Chinese Academy of Sciences, Beijing 100190, China*

The transmission electron microscopy observations of the charge ordering (CO) which governs the electronic polarization in $LuFe_2O_{4-\delta}$ clearly show the presence of a remarkable phase separation at low temperatures. Two CO ground states are found to adopt the charge modulations of $\mathbf{Q_1} = (1/3, 1/3, 0)$ and $\mathbf{Q_2} = (1/3 + \varepsilon, 1/3 + \varepsilon, 3/2)$, respectively. Our structural study demonstrates that the incommensurately $\mathbf{Q_2}$-modulated state is chiefly stable in samples with relatively lower oxygen contents. Data from theoretical simulations of the diffraction suggest that both $\mathbf{Q_1}$- and $\mathbf{Q_2}$-modulated phases have ferroelectric ordering. The effects of oxygen concentration on the phase separation and electric polarization in this layered system are discussed.

Materials that are simultaneously ferroelectric and ferromagnetic are gaining more and more attention within the scientific community. Such magnetoelectric multiferroics are examples of the broadening class of multifunctional or smart materials that combine several useful properties in the same substance. Recently, extensive investigations of the charge-frustrated $RFe_2O_{4-\delta}$ system (R=Y, Er, Yb, Tm and Lu) have revealed a rich variety of important properties [1-4], such as ferroelectricity, magnetodielectric effects, and spin/charge orders [3-8]. However, understanding the CO behavior and its dynamic nature in this layered material remains a challenge, e.g the giant magnetocapacitance effect can be understood by supposing that the charge fluctuation of $LuFe_2O_4$ arises from a visible decrease of the charge frustration under an external magnetic field. However, it is unclear by what mechanism this happens [4]. The most feature observed in $LuFe_2O_{4-\delta}$ is the visible ferroelectric polarization resulting from the ordered arrangement of $Fe^{2+}$ and $Fe^{3+}$ ions within the Fe/O double layers [3, 6-9]. Despite considerable progress in revealing the CO behaviors in this layered system, the theoretical identification and experimental observations of the CO modulations and their dominant responsibility for polarization remain controversial [1-2, 7, 9]. For instance, theoretical calculations suggest an antiferroelectric ground state [1, 2], although pyroelectric current measurements suggested the ferroelectric nature of this compound [3]. In our recent in-situ transmission electron microscopy (TEM) study, we found that $LuFe_2O_{4-\delta}$ samples commonly undergo a clear separation of different CO phases at low temperatures. This evidence shows that multiple CO states with distinctive electronic polarizations can coexist in this charge frustrated system, and these phenomena could be essentially important in establishing the mechanism by which the coupling of ferroelectricity and magnetism can be understand. In this paper, we present the first direct evidence, to our knowledge, that the CO modulations in $LuFe_2O_{4-\delta}$ adopt two typical ordered states, i.e. $\mathbf{Q_1} = (1/3, 1/3, 0)$ and $\mathbf{Q_2} = (1/3 + \varepsilon, 1/3 + \varepsilon, 3/2)$, that both result in ferroelectric polarizations. Structural phase transitions in association with phase separation have been carefully analyzed in detail.

Both polycrystalline and single crystal samples were used in present study.

Polycrystalline samples were synthesized by the conventional solid state reaction as reported in our previous publications [7]: $LuFe_2O_{4-\delta}$ samples with different oxygen content were sintered at 1473 K for 48 hrs under an oxygen partial pressure atmosphere precisely controlled by tuning the $CO_2/H_2$ ratio between 1.7 and 7. The crystal structure, ferroelectricity and magnetic properties of all $LuFe_2O_{4-\delta}$ samples have been well characterized by a variety of experimental measurements. The specimens for TEM observations were prepared by mechanical polishing to a thickness of around 20 μm, followed by ion milling. The electron-energy-loss spectroscopy (EELS) measurements and structural analysis were performed on a Tecnai F20 transmission electron microscope equipped with a post column Gatan imaging filter. The energy resolution in the EELS spectra is 0.9 eV under normal operation conditions. Under the parallel illumination, the convergence angle is about 0.7 mrad and the spectrometer collection angle is ~1.0 mrad.

It has been commonly noted in previous studies that the physical properties of the $RFe_2O_{4-\delta}$ (R=Lu, Y, Ho) materials are rather sensitive to the presence of oxygen vacancies [10-12]. Actually, oxygen deficiency (δ) in $RFe_2O_{4-\delta}$ can directly affect the $Fe^{2+}/Fe^{3+}$ ratio and, therefore, results in modification of the nature of the CO as well. In order to reveal the effects of oxygen deficiency, we systematically examined the structural properties of a series of samples in which the oxygen content is tuned by the $CO_2/H_2$ ratio during sample preparation. Figures 1a and b show the [1-10] zone-axis electron diffraction patterns taken at room temperature, illustrating visible differences in features of superstructure reflections in two samples prepared under different conditions of $CO_2/H_2 = 1.9$ or 7. The CO states shown in these diffraction patterns exhibit visible 2-dimensional features, most notably that the diffuse satellite streaks originating from the essential charge frustration reveal a clear tendency to condense into discrete satellite spots in the sample prepared with $CO_2/H_2 = 1.9$. Our further in-situ TEM observations show that the low-temperature CO states are also notably influenced by introduction of oxygen deficiencies in this layered material; and phase separation from the charge disorder toward two different charge-modulated states are often observed at lower temperatures in samples prepared with the $CO_2/H_2$ ratio between 2.0 and 7. In our recent investigations, we have made numinous attempts to quantify the oxygen contents in a series of well-characterized samples, the experimental measurements suggest that the EELS measurements could directly reveal the relative changes of oxygen concentration in $RFe_2O_{4-\delta}$ crystals. Figure 1c and 1d show the EELS data for the oxygen $K$-edge and for the Fe-$L_3$ edge in these two typical samples, illustrating the clear difference in spectral features. From the integrated spectral intensity of O $K$ and Fe $L$ edges, we can estimate the atomic ratio of O/Fe in the samples. In present case, it is roughly estimated that these two samples has a 1/30 difference in the oxygen concentration, the samples with $CO_2/H_2$ of 7 has slightly higher oxygen concentration. Though it is rather difficult to quantitatively determine small changes in oxygen concentration because of uncertainties in the partial ionization cross-sections for EELS analysis, we found that the fine structures in the O-$K$ and Fe-$L$ edges are sensitive to the chemical bonding and oxygen concentration in the $LuFe_2O_{4-\delta}$ materials.

We now turn to an in-situ cooling examination of the CO modulations in the well-characterized samples prepared with $CO_2/H_2 \approx 7$. Our in-situ cooling TEM observations revealed that the satellite streaks at (1/3, 1/3, L), shown in Fig. 1a, gain progressively in intensity with lowering temperature, and broad maxima often appear at integer L positions in the diffraction patterns. The width of CO reflection at 100K yields correlations in the c-axis direction of 3–5 double layers [7-8]. Figures 2a and b show electron diffraction patterns with clear superstructure spots taken respectively along the [1 -1 0] and [1 2 0] zone-axis directions, both at around 20 K. It can be seen that satellite spots appear at the systematic G = g ± (1/3, 1/3, 0) positions, where g is any reciprocal lattice vector with g = (H, H, 3n; n = integer). In addition, a long-periodic modulation with **q**=**Q₁**/10 + (0 0 3/2) often appears at low temperature as discussed in previous literature [7]; this modulation (**q**) is considered to be related to a certain kind of antiphase domain structures [6-7]. Figure 2b shows the diffraction pattern taken along [120] zone-axis direction, this pattern can be obtained from Fig. 2a by a 120° rotation around the c*-axis. It is commonly noted in our TEM observations that the superstructure reflections observed in many areas are not very sharp but show a visible streaking feature along the c*-axis direction, this fact suggests a short-coherent nature within the c-axis direction (~5nm) and long-coherent nature (>30nm) within the a-b plane. A clear view of the ordered state corresponding with $Q_1$ - modulation was obtained along the [1-10] zone axis direction, Figure 2c shows a high-resolution TEM image corresponding to Fig. 2a by using the (00L) spots and their (1/3, 1/3, L) satellite spots; despite the notable disordered areas, we can clearly see that the CO modulations stack regularly along the c-axis direction and yield a well-defined layered pattern.

Numerous experimental investigations are concerned with local polarization in association with the low-temperature CO in charge frustrated $LuFe_2O_{4-\delta}$ materials. For instance, (i) Mössbauer spectroscopy measurements imply a bimodal valence distribution supporting the 'order by fluctuation' mechanism for CO development [5]; (ii) theoretical calculations suggest that each Fe/O double layer in principle can be polarized as proposed by Yamada *et al.* in which the two triangular sheets in an Fe/O double layer do not have the same number of $Fe^{2+}$ and $Fe^{3+}$ ions, but instead $[Fe^{2+}]:[Fe^{3+}]=1:2$ in one triangular sheet and 2:1 in another triangular sheet. As a result, this model gives rise to an electric dipole within a $3^{1/2}a \times 3^{1/2}a$ supercell based on charge disproportion between two triangular sheets; see ref. 3 for details. Moreover, it is worth noting the appearance of charge disproportion and polarization, which together break the charge frustration degeneracy and yield the Fe-valance ratio of 2.67:2.33 between the two triangular sheets in each Fe/O double layer.

Figure 2d shows a schematic CO model illustrating a possible polarized state corresponding to the $Q_1$-modulation. It can be seen that the electric dipoles, as indicated by arrows, are actually crystallized in a ferroelectric phase, this ferroelectric layered phase reveals a notable electric polarity along the c-axis direction. Based on the local structural distortions in the Fe/O double layer [1, 13], theoretical simulation of this CO pattern could give rise to a diffraction patterns in good agreement with the experimental one (Fig.2a). Considering commensurability of the $Q_1$-modulation relative to the basic crystal structure, we can actually obtain a well-defined 3-dimensional CO superstructure with a supercell of $3^{1/2}a \times 3^{1/2}a \times c$. It is convenient to perform a ±120° rotation for this superstructure model around the c-axis to get the CO pattern in the crystallographically equivalent directions, i.e. the [1 -2 0] and [2 -1 0]-zone axis directions. As a result, we can obtain a CO pattern as shown in Fig.2e, which gives rise to a diffraction pattern consistent with Fig. 2b. In our previous studies, we interpreted $Q_1$ modulation by a charge stripe model of …. $Fe^{2+}Fe^{2.5+}Fe^{3+}Fe^{2+}Fe^{2.5+}Fe^{3+}$… in which the charge frustration was considered [7-8]. In structural point of view, this charge stripe model can used to account for the most observed structural features at low temperatures. Nevertheless, it is noted that the optical measurements and Mössbauer spectroscopy investigations [5] suggest the bimodal valence distribution in the CO ground state, where no $Fe^{2.5+}$ state is visibly determined at low temperatures. In order to facilitate the discussions on the low temperature CO behaviors, in present study, we therefore schematically illustrated all observed CO modulations based on the bimodal valence distribution.

TEM observations of the oxygen-deficient samples often reveal complex CO modulations with visible incommensurability at low temperatures. Figures 3 a-c show the electron diffraction patterns taken at ~20K for another CO phase in the $LuFe_2O_{4-\delta}$ materials sintered in an oxygen atmosphere of $CO_2/H_2 \approx 1.9$ (so called **$Q_2$**-modulated phase). This modulated structure in general shows the notable satellite spots within the (110)* reciprocal plane, the modulation wave vector with a small incommensurability ($\varepsilon$) can be written as **$Q_2$**= (1/3+$\varepsilon$, 1/3+$\varepsilon$, 3/2), where $\varepsilon$ ranges from 0 to 0.005 and can change slightly from one area to another.

The presence of a single modulation wave vector in Fig. 3a is clear evidence that the **$Q_2$** structural modulation within the Fe/O layer is one-dimensional. Sometimes three or two sets of satellite reflections, as shown in Fig. 3b and 3c, become visible in one area due to the presence of twin domains. Such twinning is expected, since the triangular lattice provides no unique orientation for the CO modulation. Considering the crystal symmetry for average structure of $LuFe_2O_{4-\delta}$ with the space group R-3m, we can get the other two symmetry-equivalent variants with the propagation vectors of **$Q_2′$** = (-2/3, 1/3, 3/2) and **$Q_2″$** = (1/3, -2/3, 3/2), where the small incommensurability ($\varepsilon$) is ignored. For instance, the satellite spots in Fig. 3b and c appear at the systematic positions of (1/3, 1/3, 5/2) and (1/3, 1/3, 1/2) can be written as: (1/3, 1/3, 5/2) =(011) +(1/3, -2/3, 3/2), arising from a $Q_2'$ variant; and (1/3, 1/3, 1/2)=(0 1 -1) + (1/3, -2/3, 3/2), arising from a $Q_2''$ varant .

To determine the local structural features in the $Q_2$-modulated phase, especially the dipole ordered state respect to structural layers, we also obtained a high-resolution TEM image at low temperature by using the $Q_2$ satellite spots. Figure 3d demonstrates a TEM image illustrating a projection of a $Q_2$-modulated area along the [1-10] zone axis direction. Despite plenty of defects, one can observe this modulation showing apparent tendency to form a body-centered stacking along the c-axis. Such an arrangement minimizes the Coulomb repulsion between the charges [14]. It is recognizable that the modulation contrast could yield a superlattice with a unit cell of $3^{1/2}a \times 3^{1/2}a \times 2d_{003}$, as clearly illustrated in Fig.3d. Figures 3e and 3f show two possible $Fe^{2+}/Fe^{3+}$ ionic models with the clear polarization within the Fe/O double layers. The uniform appearance of these two types of stacking orders results in a diffraction pattern as shown in Fig. 3a [15]. Moreover, uneven occupations of these two ordered states can explain the incommensurability observed in the low-temperature CO modulation.

On the other hand, the CO mechanism in strongly correlated systems suggests the charge ordering in general cannot simply be interpreted by the ionic orders with different valence states; an intermediate state combining both site-centered and bond-centered ordering has been demonstrated to be the energetically favored state in many systems [16, 17]. The CO in this kind of materials can intimately couple with the magnetic ordering, which allows the simultaneous development of ferroelectricity and magnetic order. A kind of superposition of these two different charge-ordering pattern can be stable as demonstrated in the $La(Ca)Mn_2O_3$ and show local dipole moments that add up to the macroscopic ferroelectric polarization. Based on this CO mechanism, we show a simple CO model in Fig. 3g with site-centered and bond-centered CO layers alternatively stacked along the c direction, this CO order pattern could give rise to a diffraction pattern in good agreement with Fig. 3a. Because the CO in the $LuFe_2O_4$ system is intimately coupled with the magnetic ordering, a theoretical analysis on the stability of the possible site-centered and bond-centered ordered states is still in progress [13].

Furthermore, it should be noted that theoretical analysis by first-principles density-functional theory based on the X-ray/neutron diffraction results suggests possible antiferroelectric ground states for $LuFe_2O_4$ [1,2], the first clear antiferroelectric configuration ground state was proposed based on neutron diffraction result very recently [1]. In order to facilitate direct comparisons of our TEM data with X-ray and neutron diffraction results, we performed a theoretical simulation for the antiferroelectric order as proposed in ref.1 and ref. 2. Figures 4a and b shows respectively the CO model for the antiferroelectric state [1] and the simulated diffraction pattern along the [1-10] zone axis, it is noted that the satellite spots appearing in this pattern are very similar with what observed in fig.3a. In addition, we also carried out our theoretical analysis on the CO mode proposed in ref. 2, figures 4c and 4d show the antiferroelectric model and the simulated diffraction pattern along the [1-10] zone axis direction. It is recognizable that these simulated data exhibit satellite spots very similar with the electron diffraction pattern shown in Fig. 3b, which have been well addressed as the coexistence of CO twinning of three symmetry-equivalent variants ($Q_2$, $Q_2'$, $Q_2''$) as discussed above. Moreover, one can also see that these two models often shows an additional set of weak superstructure spots at the systematic (0, 0, 3/2) position as indicated by an arrow in Fig. 4b and 4d; those spots never appear in the TEM observations of $LuFe_2O_{4-\delta}$ samples. Hence, we can conclude that the $LuFe_2O_{4-\delta}$ materials are very likely to have two possible types of CO ground states; they both could have the ferroelectric arrangements of polarized Fe/O double layers.

We now proceed to discuss the effects of oxygen vacancies on the CO nature in the $LuFe_2O_{4-\delta}$ materials. Actually, lattice defects in correlation with oxygen deficiencies in this layered material are considered to be important in understanding the material's dielectric properties and ferrimagnetism [10-12]. Our in-situ TEM experimental investigations also suggest that the particular $LuFe_2O_{4-\delta}$ crystals prepared in an oxygen atmosphere with a $CO_2/H_2$ ratio between 1.9 and 7 undergo a clear phase separation recognizable as transitions of charge disorder, via the intermediate two-dimensional CO, developing toward the coexistence of $\mathbf{Q_1}$ and $\mathbf{Q_2}$ phases as temperature is lowered. As a result, the low-temperature diffraction patterns commonly exhibit much more complex patterns of superstructure spots from small domains with different CO variants. Figure 5a shows an electron diffraction pattern taken from an area with the coexistence of $\mathbf{Q_1}$- and $\mathbf{Q_2}$- phases. Figure 5b shows a dark-field TEM image taken at 100K using satellite reflections. In order to better depict the structural properties of the charge ordered state, we have smoothed the image and created a two-level contour map and illustrate the complex domain-like contrast for the CO state. The bright regions represent domains that contribute to the $\mathbf{Q_1}$- reflections.

In conclusion, we show by in-situ TEM observations that charge frustrated $LuFe_2O_{4-\delta}$ in general undergoes a clear phase separation with decreasing temperature and yields two different CO ground states, i.e. $\mathbf{Q_1}$ = (1/3, 1/3, 0) and $\mathbf{Q_2}$ = (1/3 + $\varepsilon$, 1/3+ $\varepsilon$, 3/2). Our experimental results demonstrate that the oxygen deficient samples mainly adopt the incommensurate $\mathbf{Q_2}$-modulated state. Theoretical simulations of these two different types of CO states show that they both exhibit ferroelectric polarization. Coexistence and competition of these two CO phases are also observed in many $LuFe_2O_{4-\delta}$ materials depending on oxygen content and temperature. It is also noted that the $LuFe_2O_{4-\delta}$ materials often have much greater spontaneous electronic polarization than that of the recently discovered multiferroic materials [18]. Hence, the multiferroic $LuFe_2O_{4-\delta}$ with a specific CO state can be a good example for envisaging a coupling between ferroelectric and magnetic properties that lead to magnetic-field-switchable ferroelectricity or vice versa.

**Acknowledgements** This work is supported by the National Science Foundation of China, the Knowledge Innovation Project of the Chinese Academy of Sciences, and the 973 projects of the Ministry of Science and Technology of China.

**Figure Captions**

**Figure 1 Superstructure modulations in two LuFe$_2$O$_{4-\delta}$ samples prepared under different conditions.** Electron diffraction patterns taken along the [1-10] zone-axis direction at room temperature from samples prepared with (a) CO$_2$/H$_2$= 7 and (b) CO$_2$/H$_2$=1.9, respectively, illustrating the difference in satellite spots features. Electron energy loss spectra of (c) oxygen *K*-edges and (d) Fe-*L$_3$* edges for these two samples (CO$_2$/H$_2$= 7 sample is marked as a, and CO$_2$/H$_2$= 1.9 is marked as b) illustrating the different structural features, the oxygen concentration and average valence state of Fe ions in the sample prepared with CO$_2$/H$_2$= 7 are relatively higher.

**Figure 2 Charge ordering and spontaneous polarization in Q$_1$-modulated phase.** Electron diffraction patterns observed at about 20 K taken along the (a) [1-10] and (b) [120] zone-axis directions respectively. (c) [1-10] HRTEM image using the **Q$_1$** - satellite spots showing the CO modulation stacking regularly along the c-axis direction. (d) Schematic CO model of the [1-10] projection illustrating the possible polarization as indicated by arrows. (e) Schematic CO model of the [1 2 0] zone projection.

**Figure 3 Charge ordering and spontaneous polarization in Q$_2$ modulated LuFe$_2$O$_{4-\delta}$ phase.** Electron-diffraction patterns taken along [1-10] zone-axis directions at ~20K from samples (a) contains a single modulation vector, two (b) and three (c) sets of satellite reflections due to the presence of twin domains. (d) HREM image demonstrates a projection of a **Q₂**-modulated area along the [1-10] axis direction showing an obvious tendency to form a body-centered stacking along the c-axis. A $3^{1/2}a \times 3^{1/2}a \times 2d_{003}$ supercell is clearly illustrated. (e)-(f) Two possible $Fe^{2+}/Fe^{3+}$ ionic models with the clear ferroelectric polarized state. (g) A CO model illustrating the site-centered and bond-centered CO layers alternatively stacking along the c-axis direction.

**Figure 4 Structural models for the CO states and simulated diffraction patterns.** (a) Antiferroelectric model proposed in ref. 1. (b) Simulated diffraction pattern along [1-10] zone axis direction. (c) CO model proposed in ref. 2. (d) Simulated diffraction pattern along [1-10] zone axis direction, illustrating the additional satellite spots.

**Figure 5 Phase separation in $LuFe_2O_{4-\delta}$.** (a) An electron diffraction pattern taken from an area with the coexistence of **Q₁**- and **Q₂**-phases. (b) A two-level contour map from a dark-field TEM image of the **Q₁**- phase taken at 100K emphasizing the domain structures within the examined area.

Fig1

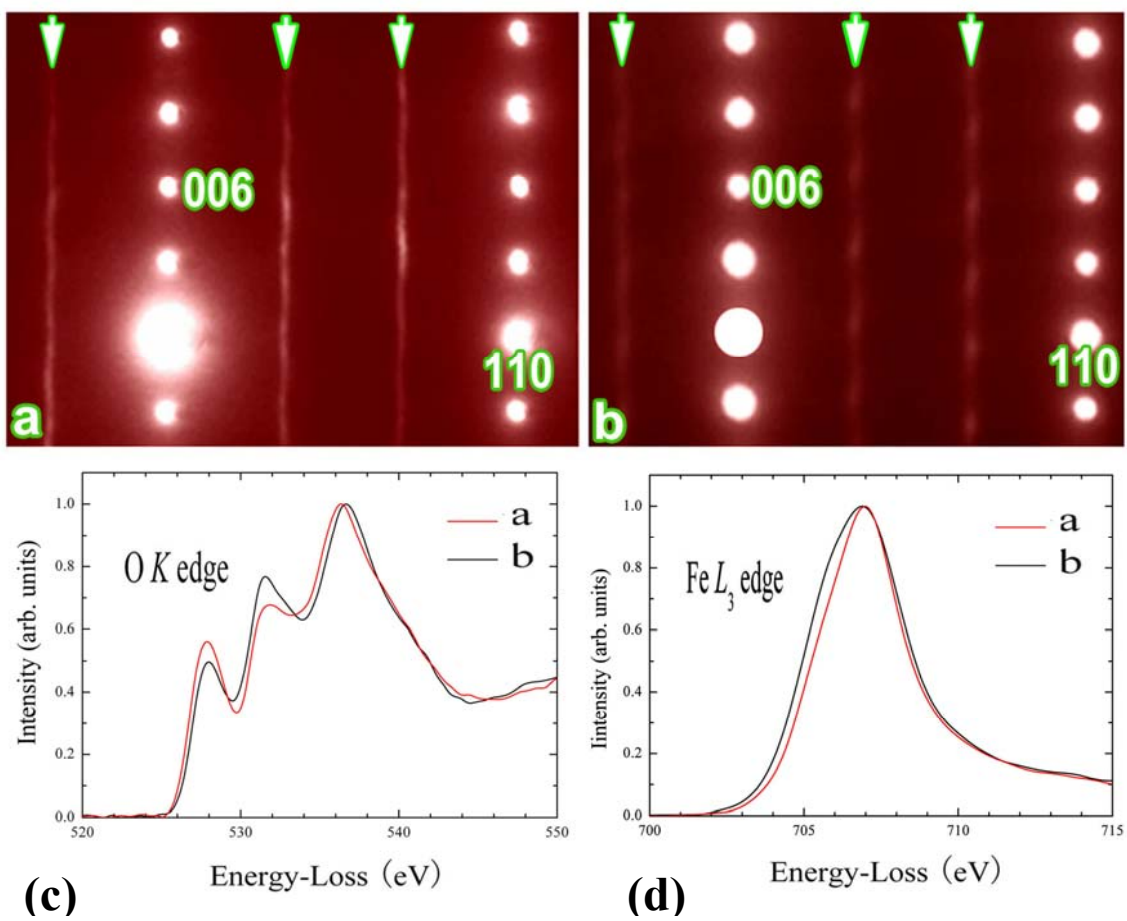

Fig2

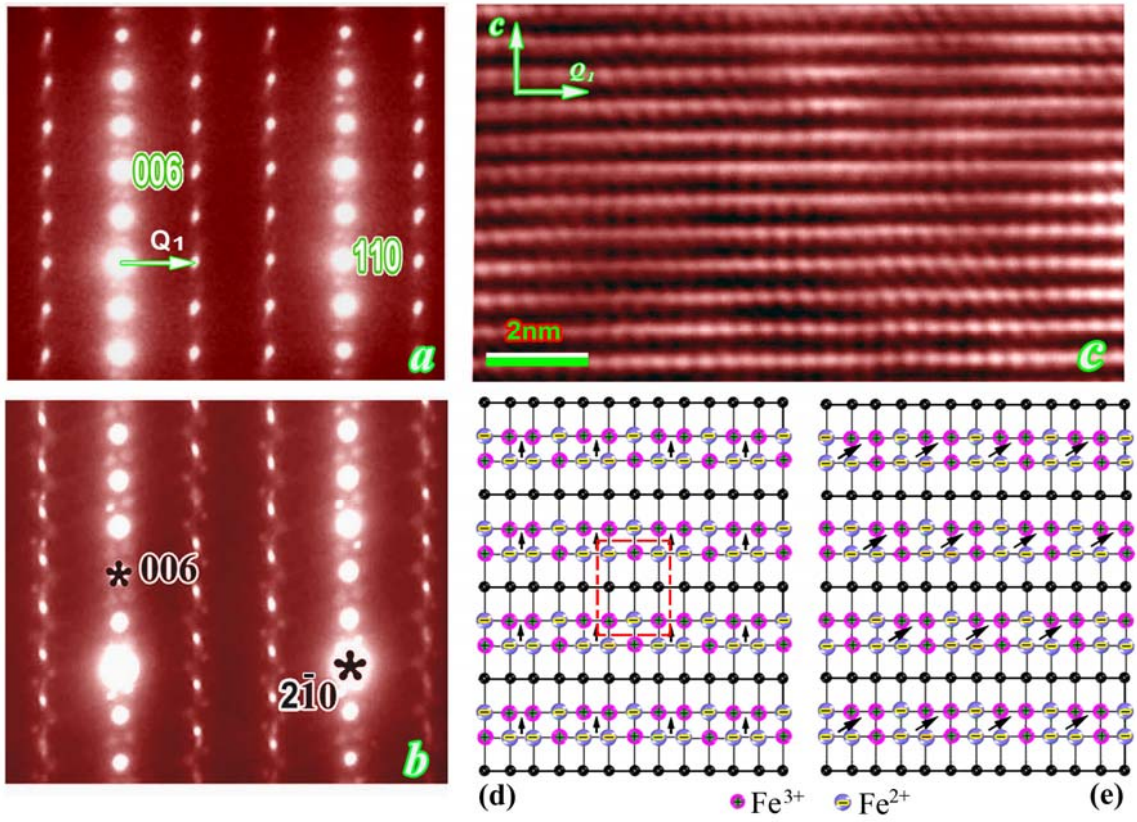

Fig3

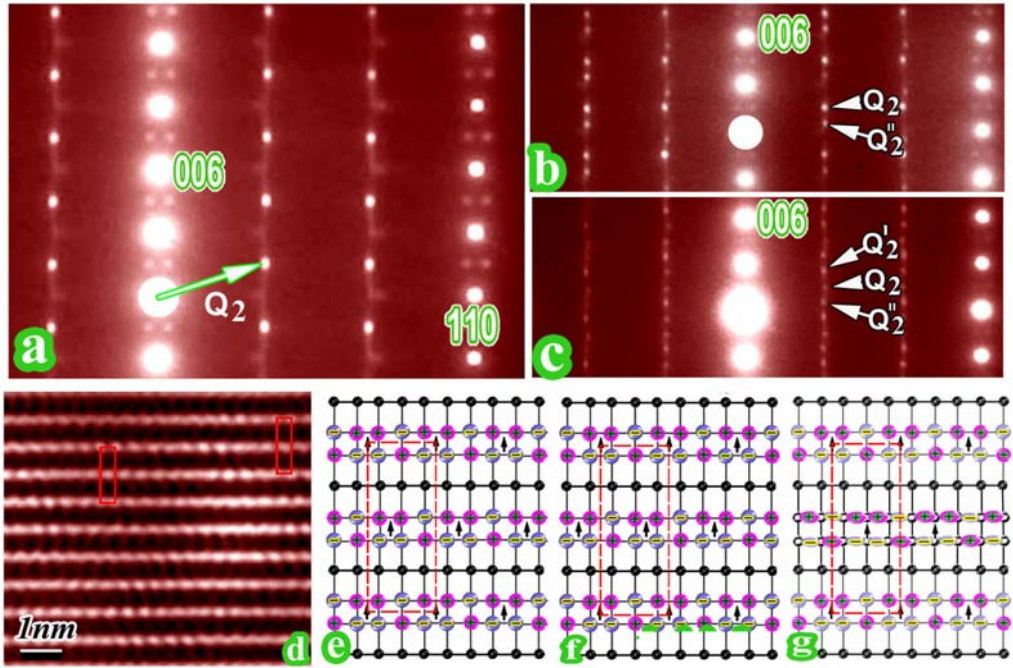

Fig4

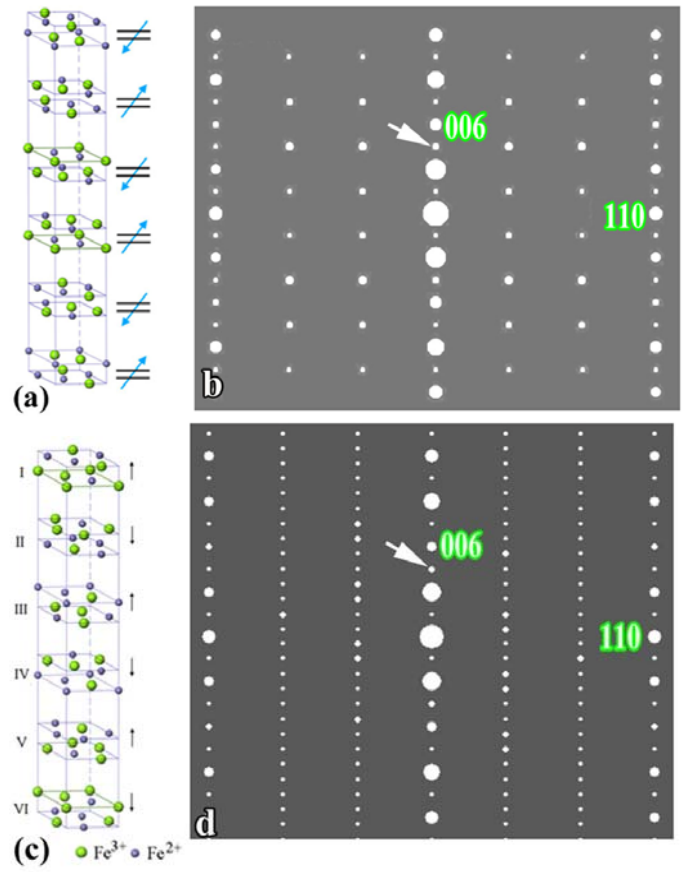

Fig5

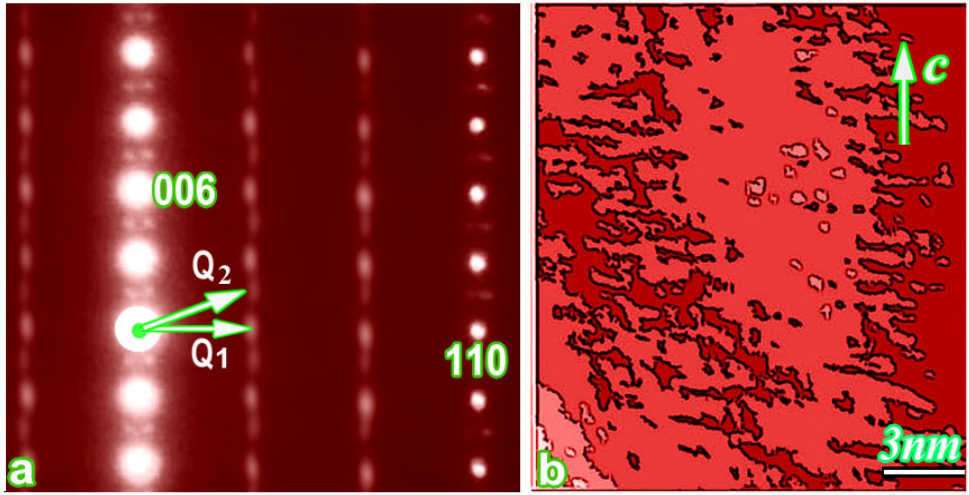